\documentclass[prl,twocolumn,showpacs,superscriptaddress]{revtex4}
\usepackage{graphicx,amsmath}

\newcommand{\figwidth}{1.00\columnwidth}
\newcommand{\eq}[1]{Eq.(\ref{#1})}
\newcommand{\fig}[1]{Fig.~\ref{#1}}
\newcommand{\avg}[1]{\langle #1 \rangle}
\newcommand{\savg}[1]{[ #1 ]_{\rm av}}
\newcommand{\olcite}[1]{Ref.~\onlinecite{#1}}

\newcommand{\sigmac}{\sigma_{\rm c}}
\newcommand{\sigmap}{\sigma_{\rm p}}
\newcommand{\sigmam}{\sigma_{\rm M}}
\newcommand{\nc}{N_{\rm c}}
\newcommand{\zc}{z_{\rm c}}
\newcommand{\zp}{z_{\rm p}}

\newcommand{\etam}{\eta_{\rm M}}
\newcommand{\etac}{\eta_{\rm c} }
\newcommand{\etaccr}{\eta_{\rm c,cr} }
\newcommand{\etapr}{\eta_{\rm p}^{\rm r}}
\newcommand{\etaprcr}{\eta_{\rm p,cr}^{\rm r}}
\newcommand{\pc}{P_L}
\newcommand{\kb}{k_{\rm B}}

\begin{document}

\title{Critical behavior of colloid-polymer mixtures in random porous media}

\author{R. L. C. Vink}
\affiliation{Institut f\"ur Theoretische Physik, Heinrich-Heine-Universit\"at 
D\"usseldorf, Universit\"atsstra{\ss}e 1, 40225
D\"usseldorf, Germany}

\author {K. Binder}
\affiliation{Institut f\"ur Physik, Johannes Gutenberg-Universit\"at 
Mainz, Staudinger Weg 7, 55099 Mainz, Germany}

\author{H. L\"owen}
\affiliation{Institut f\"ur Theoretische Physik, Heinrich-Heine-Universit\"at 
D\"usseldorf, Universit\"atsstra{\ss}e 1, 40225
D\"usseldorf, Germany}

\date{\today}

\begin{abstract}

We show that the critical behavior of a colloid-polymer mixture inside a 
random porous matrix of quenched hard spheres belongs to the universality 
class of the random-field Ising model. We also demonstrate that 
random-field effects in colloid-polymer mixtures are surprisingly strong. 
This makes these systems attractive candidates to study random-field 
behavior experimentally.

\end{abstract}


\maketitle

One longstanding problem in the phase behavior of fluids concerns the 
universality class of the liquid-gas transition in random porous media. 
Conceivable universality classes are those of the pure Ising model, the 
(bond or site) diluted Ising model \cite{berche.chatelain.ea:2004}, or the 
random-field Ising model \cite{imbrie:1984, imry.ma:1975, 
nattermann:1998}. It was suggested by de~Gennes \cite{gennes:1984} that 
the universality class is that of the random-field Ising model. However, 
resolving the universality class experimentally is difficult, 
partly because well-characterized porous media are scarce. The prototype 
realization is silica aerogel, which has the disadvantage that the 
coupling between the porous medium and the fluid is weak. This is evident 
from the small shift in the critical temperature observed in these systems 
\cite{wong.chan:1990, wong.kim.ea:1993}
\begin{equation}\label{eq:one}
  \delta \equiv |T_M - T_P|/T_P \approx 6 \times 10^{-3},
\end{equation}
where $T_P$ is the critical temperature in the pure system, and $T_M$ the
critical temperature in the aerogel matrix. The universality class, 
consequently, could not be resolved in these experiments.

In theoretical approaches, the porous medium is usually modeled by an 
equilibrium configuration of fixed spheres. The fluid particles are then 
allowed to migrate through the medium. Interestingly, these 
quenched-annealed systems display pronounced shifts in the critical 
temperature, typically $\delta > 0.2$ \cite{alvarez.levesque.ea:1999, 
sarkisov.monson:2000, grandis.gallo.ea:2004, 
scholl-paschinger.levesque.ea:2001, schmidt.scholl-paschinger.ea:2002, 
kierlik.rosinberg.ea:1996}. Compared to aerogel, the coupling between the 
fluid and the porous medium is thus much stronger, making 
quenched-annealed fluids attractive model systems.

Nevertheless, the universality class of quenched-annealed systems has not 
yet been determined. The most direct approach, which is to measure the 
critical exponents, is cumbersome because the critical exponents of the 
pure and diluted Ising model are rather similar, whereas for the 
random-field Ising model (RFIM), the critical exponents are not precisely 
known in any case. However, there is an additional feature in the critical 
behavior which can be used to resolve the universality class. A striking 
feature of the RFIM is that the standard hyperscaling relation between 
critical exponents is violated, and replaced by the modified relation 
$\gamma+2\beta=\nu(d-\theta)$ \cite{villain:1982, fisher:1986}. Here, 
$d=3$ is the spatial dimension; $\beta$, $\gamma$, and $\nu$ are the 
critical exponents of the order parameter, susceptibility, and correlation 
length, respectively; while $\theta$ is a third (probably not independent 
\cite{gofman.adler.ea:1993}) exponent called the ``violation of 
hyperscaling'' exponent. For the pure ($\beta \approx 0.326; \gamma 
\approx 1.239; \nu \approx 0.630$ \cite{fisher.zinn:1998}) and diluted 
($\beta \approx 0.35; \gamma \approx 1.34; \nu \approx 0.68$ 
\cite{berche.chatelain.ea:2004}) Ising models hyperscaling is {\it not} 
violated, implying $\theta=0$. In contrast, for the RFIM ($\beta < 0.13; 
\gamma \approx 1.7-1.9; \nu \approx 1.02-1.1$ \cite{rieger:1995, 
newman.barkema:1996}) one has $\theta \sim 1$.

The aim of this Letter is two-fold. First, we demonstrate that 
hyperscaling is violated in quenched-annealed systems, using large-scale 
computer simulations and finite-size scaling. This fixes the critical 
behavior of quenched-annealed systems into the universality class of the 
RFIM, confirming the conjecture of de~Gennes. Secondly, we show that 
quenched-annealed systems also pave the way toward exciting new 
experiments. As we will discuss, the considered quenched-annealed system 
is realized experimentally in a colloid-polymer mixture. Compared to 
aerogel, random-field effects should become easier to detect, due to the 
strong coupling between the fluid and the porous medium. 

The outline of this Letter is as follows. First, we introduce our 
quenched-annealed model. Next, simulations are used to show that 
hyperscaling is violated. In addition, the critical exponents extracted 
from our data are shown to be compatible with those of the RFIM. Finally, 
we discuss how a quenched-annealed system can be realized experimentally 
in a colloid-polymer mixture.

We consider colloid-polymer mixtures within the framework of the 
Asakura-Oosawa-Vrij (AOV) model \cite{asakura.oosawa:1954, vrij:1976}. The 
AOV model is known to reproduce experimental observations remarkably well, 
including bulk phase-separation \cite{aarts.tuinier.ea:2002}, interfacial 
properties \cite{brader.evans.ea:2002}, and gelation 
\cite{bergenholtz.poon.ea:2003}. In this model, colloids and polymers are 
treated as effective spheres of diameter $\sigmac$ and $\sigmap$, 
respectively. The colloid-to-polymer size ratio is denoted as $q \equiv 
\sigmap / \sigmac$. Hard sphere interactions are assumed between 
colloid-colloid and colloid-polymer pairs, while polymer-polymer pairs can 
interpenetrate freely. The simulations are performed in the grand 
canonical ensemble, where the volume $V$ and the respective 
(dimensionless) fugacities, $\zc$ and $\zp$, of colloids and polymers are 
fixed, while the number of particles inside $V$ fluctuates (lengths are 
expressed in units of $\sigmac$). Following convention, the polymer 
fugacity is expressed by the polymer reservoir packing fraction $\etapr = 
\pi \zp q^3 / 6$. The colloid packing fraction is $\etac = \pi \sigmac^3 
\nc / (6V)$, with $\nc$ the number of colloids in the system. In the 
absence of quenched disorder, the phase behavior of the AOV model is well 
understood. For sufficiently large $q$, the AOV model phase separates into 
a colloid-rich phase (the colloidal liquid) and colloid-poor phase (the 
colloidal vapor), if $\etapr$ exceeds a critical value $\etaprcr$ 
\cite{lekkerkerker.poon.ea:1992}. The binodal exhibits an Ising critical 
point \cite{vink.horbach:2004*1}. The phase transition is driven by 
$\etapr$, which thus plays a role similar to that of inverse temperature 
in liquid-vapor transitions of simple fluids.

We now consider the AOV model inside a random porous medium. The medium is 
modeled as an equilibrium ideal-gas configuration of spheres of diameter 
$\sigmam = \sigmac$ at fixed packing fraction $\etam=0.05$ (in the 
terminology of \olcite{physrevb.48.3095}, this resembles anticorrelated 
disorder). We set $q=1.0$, assuming hard-sphere interactions between 
colloid-matrix pairs, while the polymer-matrix interaction is left ideal. 
We use a cubic simulation box of edge $L$ with periodic boundary 
conditions. We aim to measure the order parameter distribution 
$\savg{\pc}$. Here, $\pc \equiv P_L(\etac)$ is the probability of 
observing a system with colloid packing fraction $\etac$ measured for one 
realization of the matrix, while $\savg{\cdot}$ denotes an average over 
different matrix realizations. For each random matrix, a grand canonical 
Monte-Carlo simulation of the AOV model is performed, using a cluster move 
\cite{vink.horbach:2004*1}. Colloids and polymers are inserted and removed 
from the simulation box at random, with the constraint that 
colloid-colloid, colloid-polymer, and colloid-matrix overlaps are 
forbidden. During the simulations, the (quenched) matrix particles remain 
fixed. The number of colloids in the simulation box fluctuates, and this 
is used to measure $\savg{\Delta F (\nc,\nc+1)}$, defined as the free 
energy difference between the state with $\nc$ and $\nc+1$ colloids, 
averaged over typically 250 matrix realizations. By successively measuring 
the free energy difference, the total averaged free energy $\savg{W}$ as 
function of $\etac$ results \cite{virnau.muller:2004}. The latter is 
related to the sought-for distribution $\savg{\pc} \propto 
e^{-\savg{W}/\kb T}$, with $\kb$ the Boltzmann constant and $T$ the 
temperature.

\begin{figure}
\begin{center}
\includegraphics[clip=,width=\figwidth]{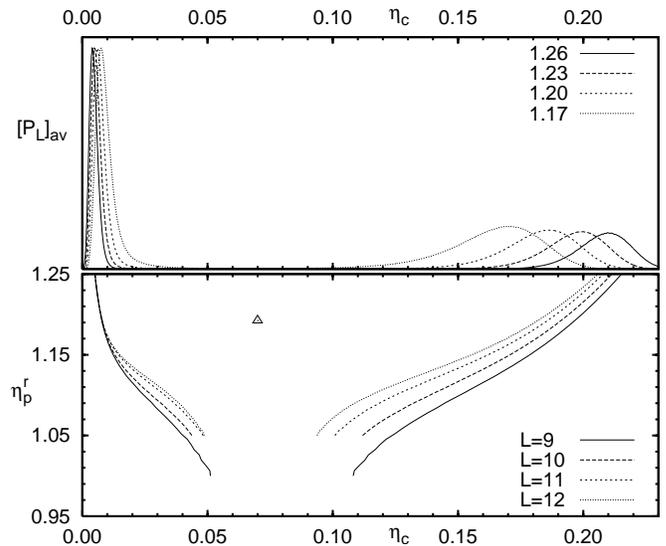}

\caption{\label{prob} {\it top frame:} Order parameter distribution 
$\savg{\pc}$ for $L=12$ and several values of $\etapr$. {\it 
lower frame:} Binodal curves, obtained by reading-off the peak positions 
in $\savg{\pc}$, for several system-sizes $L$. The triangle, at $\etaccr 
\approx 0.070$ and $\etaprcr \approx 1.192$, marks the location of the 
critical point in the thermodynamic limit.}

\end{center}
\end{figure}

\begin{figure}
\begin{center}
\includegraphics[clip=,width=\figwidth]{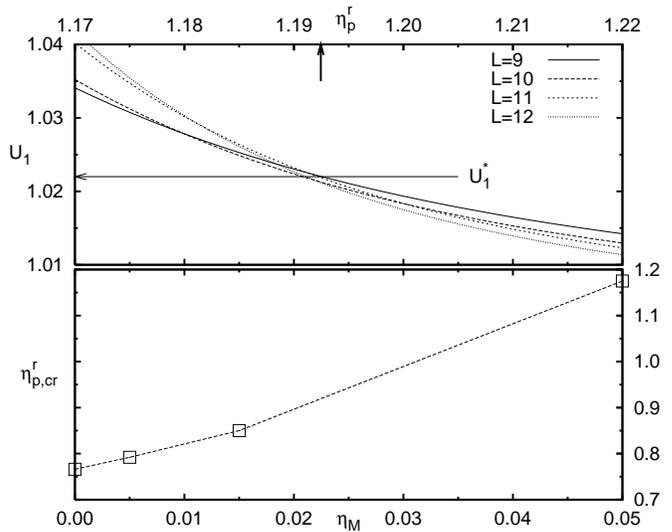}

\caption{\label{cum} {\it top frame:} Cumulant analysis of the AOV model 
inside a porous medium at $\etam=0.05$. Shown is $U_1$ versus $\etapr$ for 
several system-sizes $L$. The vertical arrow marks our best estimate of 
$\etaprcr$; the horizontal arrow defines $U_1^\star$. {\it lower frame:} 
$\etaprcr$ as function of $\etam$; the line serves to guide the eye.}

\end{center}
\end{figure}

At two-phase coexistence, $\savg{\pc}$ becomes double-peaked, where the 
peak at low (high) $\etac$ reflects the colloidal vapor (liquid) phase. 
The coexisting phase densities follow from the average peak positions. 
Typical distributions $\savg{\pc}$ are shown in the upper frame of 
\fig{prob}, for several values of $\etapr$. By recording the peak 
positions as function of $\etapr$, the binodal is found (see lower frame). 
Unusual behavior is revealed. For pure Ising critical behavior, 
well-separated peaks in the order parameter distribution indicate that one 
is well away from the critical point, and inside the two-phase region of 
the phase diagram. Finite size effects in the peak positions should then 
be small. In contrast, even though the peaks in $\savg{\pc}$ are 
well-separated for $\etapr \approx 1.17$ and higher, the binodal continues 
to display a pronounced $L$-dependence. To locate the critical point, we 
have measured the $L$-dependence of the cumulant $U_1 \equiv 
\savg{\avg{m^2}} / \savg{\avg{|m|}}^2$ with $m = |\etac - 
\savg{\avg{\etac}}|$. At the critical point, the cumulant becomes 
system-size independent \cite{binder:1981}. Plots of $U_1$ versus 
$\etapr$, for several system-sizes $L$, are expected to show a common 
intersection point, leading to an estimate of $\etaprcr$. The result is 
shown in the top frame of \fig{cum}, yielding $\etaprcr = 1.192 \pm 
0.005$. To determine the critical colloid packing fraction $\etaccr$, the 
quantity $\savg{\avg{\etac}}(L)$ evaluated at $\etaprcr$, was linearly 
extrapolated in $1/L$, yielding $\etaccr \approx 0.070$.

We thus find that the order parameter distribution at the critical point 
remains sharp, featuring two well-separated and non-overlapping peaks. 
Violation of hyperscaling then follows from the $L$-dependence of the peak 
positions and the root-mean-square peak widths \cite{eichhorn.binder:1995, 
eichhorn.binder:1996}. For the difference $\Delta$ between the liquid and 
vapor peak positions, finite size scaling predicts $\Delta \propto 
L^{-\beta/\nu}$. Similarly, for the width $\chi$ of the vapor or liquid 
peak $\chi \propto L^{(\gamma/\nu-d)/2}$. The relative peak width thus 
becomes $w_{\rm r} \equiv \chi / \Delta \propto L^\omega$ with $\omega = 
(\gamma/\nu-d)/2 + \beta/\nu$. In case hyperscaling holds $\omega=0$, and 
a finite relative width $w_{\rm r} > 0$ in the thermodynamic limit $L \to 
\infty$ is retained. In contrast, when hyperscaling is violated and 
$\omega<0$, $w_{\rm r}$ vanishes in the thermodynamic limit, leading to an 
order parameter distribution featuring two $\delta$-peaks. Substitution of 
the RFIM critical exponents indeed yields $\omega < 0$, implying $w_{\rm 
r} \to 0$ for $L \to \infty$, consistent with our observations. Moreover, 
since the critical order parameter distribution tends to a sum of two 
$\delta$-peaks, the value $U_1^\star$ of the cumulant {\it at} the 
critical point (horizontal arrow in \fig{cum}) approaches the trivial 
value $U_1=1$. Indeed, the cumulants of \fig{cum} intersect at a value 
close to one, consistent with RFIM critical behavior, but ruling out pure 
and diluted Ising critical behavior, where $U_1^\star$ is significantly 
different from unity (see also \fig{cross} below).

\begin{figure}
\begin{center}
\includegraphics[clip=,width=\figwidth]{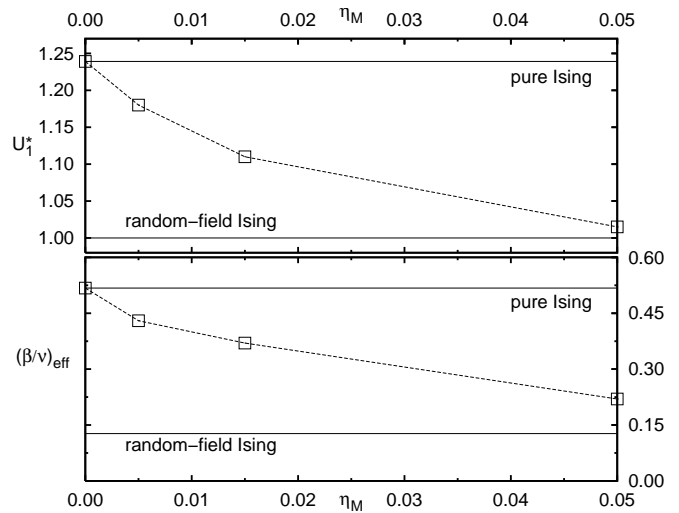}

\caption{\label{cross} Evidence of the crossover in critical behavior, 
from pure Ising toward random-field Ising, as function of $\etam$. The top 
frame shows $U_1^\star$ as function of $\etam$. The lower frame shows the 
effective critical exponent ratio $\beta/\nu$ as function of $\etam$. The 
dashed lines serve to guide the eye; horizontal lines show pure Ising 
and RFIM values.}

\end{center}
\end{figure}

Having established that quenched-annealed systems show RFIM critical 
behavior, the experimental realization of such a model will be discussed. 
To this end, the crossover in critical behavior \cite{fisher:1974, 
binder.deutsch:1992}, from pure Ising to RFIM, must be addressed. Defining 
$t$ as the relative distance from the critical point, the approach to the 
critical point, at $t=0$, is characterized by two regimes: $t<t_x$ and 
$t>t_x$, with $t_x$ the crossover temperature. For fluids in porous media, 
RFIM critical behavior is observed only when $t<t_x$. For $t>t_x$, the 
critical behavior is still dominated by the pure Ising model, and 
``effective'' critical behavior is observed instead (in this case a 
combination of RFIM and pure Ising universality). Whether RFIM critical 
behavior can be observed depends crucially on $t_x$. If $t_x$ is very 
small, precise temperature control is required, which may be difficult to 
realize experimentally. Note that $t_x$ is a non-universal quantity, 
dependent on the particle interactions, and, most importantly, on the 
packing fraction of the porous medium $\etam$. By increasing $\etam$, 
random field effects are expected to become more pronounced, implying a 
larger $t_x$. For silica aerogel, $t_x$ is clearly very small, since the 
measured critical exponents do not differ much from those of the pure 
Ising model \cite{wong.chan:1990, wong.kim.ea:1993}. In contrast, our 
results for the quenched-annealed model show pronounced RFIM behavior, 
indicating a much larger $t_x$. For experimental applications, it then 
becomes relevant to know which values of $\etam$ are required, in order to 
enable measurements in the regime $t<t_x$ and observe RFIM behavior. 

Shown in the lower frame of \fig{cum} is $\etaprcr$ as function of $\etam$ 
\footnote{For these data, we set $q=0.8$, $\sigmac=\sigmam$, and also 
introduce hard-core matrix-polymer interactions.}. Defining the analogue 
of \eq{eq:one} as $\delta \equiv |\etaprcr(M) - \etaprcr(P)| / 
\etaprcr(P)$, with $\etaprcr(M)$ the critical value of $\etapr$ in the 
presence of the random matrix, and $\etaprcr(P)$ the corresponding value 
in the pure ($\etam=0$) system, $\delta$ is found to increase from $\delta 
\approx 0.03$ ($\etam=0.005$) up to $\delta \approx 0.5$ ($\etam=0.05$). 
This confirms our expectation that, by increasing $\etam$, random-field 
effects grow stronger, and $t_x$ becomes larger. Evidence for the 
crossover in critical behavior is presented in the top frame of 
\fig{cross}. Shown is $U_1^\star$ as function of $\etam$. Recall that 
$U_1^\star$ is defined as the value of the cumulant at the critical point, 
see the top frame of \fig{cum}. The horizontal lines correspond to $U_{\rm 
1,pure}^\star \approx 1.2391$ \cite{luijten.fisher.ea:2002} of the pure 
Ising model, and the (exact) RFIM value $U_1^\star=1$. The figure 
strikingly illustrates effective critical behavior, between that of the 
pure Ising model and the RFIM, with a pronounced drift toward the latter, 
as $\etam$ increases. Additional confirmation of the crossover is obtained 
from the critical exponent ratio $\beta/\nu$. Shown in the lower frame of 
\fig{cross} is $\beta/\nu$ as function of $\etam$. The upper horizontal 
line shows the pure Ising value; the lower line is an upper bound for the 
corresponding RFIM value \cite{rieger:1995, newman.barkema:1996}. Again, a 
clear drift toward the RFIM value is observed.

\fig{cross} then provides a clear indication which value of $\etam$ to use 
in an experiment. The relative distance from the critical point that can 
be reached in simulations is nowadays $t \approx 10^{-3}$ 
\cite{kim.fisher.ea:2003}; similar precision is also achieved in 
experimental colloid-polymer systems \cite{royal.aarts.ea:2006}. Although 
this precision is rather low compared to what is achieved in atomic 
fluids, \fig{cross} nevertheless shows pronounced deviations from pure 
Ising behavior already at $\etam=0.015$, with the crossover to RFIM being 
nearly completed at $\etam=0.05$. This suggests $\etam>0.05$ as the 
optimal regime for experiments. Such packing fractions are surprisingly 
low, and easily realized in aggregated colloidal rods 
\cite{physreve.61.626} or spheres \cite{kluijtmans.philipse:1999}. There 
may even be the exciting possibility to generate the porous medium by 
optically trapping some of the colloidal particles 
\cite{vossen.van-der-horst.ea:2004}. An additional advantage is that 
colloidal particles, due to their mesoscopic size, allow for very detailed 
investigations of fluid phase behavior. By using confocal microscopy, 
individual particles can be visualized and tracked directly in real-space 
\cite{blaaderen:1997}. This has already enabled particle-level 
investigations of interface fluctuations \cite{aarts.schmidt.ea:2004} and 
bulk critical behavior \cite{royal.aarts.ea:2006}. The present results 
indicate that the experimental verification of random-field behavior is 
feasible in colloid-polymer mixtures.

In summary, we have used large-scale Monte-Carlo simulations to resolve 
the universality class of the quenched-annealed model. The universality 
class was shown to be that of the random-field Ising model, as was evident 
from the violation of hyperscaling, and the behavior of (effective) 
critical quantities. This confirms the conjecture of de~Gennes. In 
addition, we have demonstrated the potential of colloid-polymer mixtures 
in the experimental detection of random-field critical behavior, providing 
a valuable alternative over aerogel-based systems.

\acknowledgments

This work was supported by the {\it Deutsche Forschungsgemeinschaft} under 
the SFB-TR6 (project sections D3 and A5). We thank M. Schmidt for 
stimulating discussions.

\bibstyle{revtex}
\bibliography{mainz}

\begin{thebibliography}{41}
\expandafter\ifx\csname natexlab\endcsname\relax\def\natexlab#1{#1}\fi
\expandafter\ifx\csname bibnamefont\endcsname\relax
  \def\bibnamefont#1{#1}\fi
\expandafter\ifx\csname bibfnamefont\endcsname\relax
  \def\bibfnamefont#1{#1}\fi
\expandafter\ifx\csname citenamefont\endcsname\relax
  \def\citenamefont#1{#1}\fi
\expandafter\ifx\csname url\endcsname\relax
  \def\url#1{\texttt{#1}}\fi
\expandafter\ifx\csname urlprefix\endcsname\relax\def\urlprefix{URL }\fi
\providecommand{\bibinfo}[2]{#2}
\providecommand{\eprint}[2][]{\url{#2}}

\bibitem[{\citenamefont{Berche et~al.}(2004)\citenamefont{Berche, Chatelain,
  Berche, and Janke}}]{berche.chatelain.ea:2004}
\bibinfo{author}{\bibfnamefont{P.~E.} \bibnamefont{Berche}},
  \bibinfo{author}{\bibfnamefont{C.}~\bibnamefont{Chatelain}},
  \bibinfo{author}{\bibfnamefont{B.}~\bibnamefont{Berche}}, \bibnamefont{and}
  \bibinfo{author}{\bibfnamefont{W.}~\bibnamefont{Janke}},
  \bibinfo{journal}{Eur. Phys. J. B} \textbf{\bibinfo{volume}{38}},
  \bibinfo{pages}{463} (\bibinfo{year}{2004}).

\bibitem[{\citenamefont{Imbrie}(1984)}]{imbrie:1984}
\bibinfo{author}{\bibfnamefont{J.~Z.} \bibnamefont{Imbrie}},
  \bibinfo{journal}{Phys. Rev. Lett.} \textbf{\bibinfo{volume}{53}},
  \bibinfo{pages}{1747} (\bibinfo{year}{1984}).

\bibitem[{\citenamefont{Imry and Ma}(1975)}]{imry.ma:1975}
\bibinfo{author}{\bibfnamefont{Y.}~\bibnamefont{Imry}} \bibnamefont{and}
  \bibinfo{author}{\bibfnamefont{S.-k.} \bibnamefont{Ma}},
  \bibinfo{journal}{Phys. Rev. Lett.} \textbf{\bibinfo{volume}{35}},
  \bibinfo{pages}{1399} (\bibinfo{year}{1975}).

\bibitem[{\citenamefont{Nattermann}(1998)}]{nattermann:1998}
\bibinfo{author}{\bibfnamefont{T.}~\bibnamefont{Nattermann}}, in
  \emph{\bibinfo{booktitle}{Spin Glasses and Random Fields}}, edited by
  \bibinfo{editor}{\bibfnamefont{A.~P.} \bibnamefont{Young}}
  (\bibinfo{publisher}{World Scientific}, \bibinfo{address}{Singapore},
  \bibinfo{year}{1998}), p. \bibinfo{pages}{277}.

\bibitem[{\citenamefont{de~Gennes}(1984)}]{gennes:1984}
\bibinfo{author}{\bibfnamefont{P.~G.} \bibnamefont{de~Gennes}},
  \bibinfo{journal}{J. Phys. Chem.} \textbf{\bibinfo{volume}{88}},
  \bibinfo{pages}{6469} (\bibinfo{year}{1984}).

\bibitem[{\citenamefont{Wong and Chan}(1990)}]{wong.chan:1990}
\bibinfo{author}{\bibfnamefont{A.~P.~Y.} \bibnamefont{Wong}} \bibnamefont{and}
  \bibinfo{author}{\bibfnamefont{M.~H.~W.} \bibnamefont{Chan}},
  \bibinfo{journal}{Phys. Rev. Lett.} \textbf{\bibinfo{volume}{65}},
  \bibinfo{pages}{2567} (\bibinfo{year}{1990}).

\bibitem[{\citenamefont{Wong et~al.}(1993)\citenamefont{Wong, Kim, Goldburg,
  and Chan}}]{wong.kim.ea:1993}
\bibinfo{author}{\bibfnamefont{A.~P.~Y.} \bibnamefont{Wong}},
  \bibinfo{author}{\bibfnamefont{S.~B.} \bibnamefont{Kim}},
  \bibinfo{author}{\bibfnamefont{W.~I.} \bibnamefont{Goldburg}},
  \bibnamefont{and} \bibinfo{author}{\bibfnamefont{M.~H.~W.}
  \bibnamefont{Chan}}, \bibinfo{journal}{Phys. Rev. Lett.}
  \textbf{\bibinfo{volume}{70}}, \bibinfo{pages}{954} (\bibinfo{year}{1993}).

\bibitem[{\citenamefont{\'{A}lvarez et~al.}(1999)\citenamefont{\'{A}lvarez,
  Levesque, and Weis}}]{alvarez.levesque.ea:1999}
\bibinfo{author}{\bibfnamefont{M.}~\bibnamefont{\'{A}lvarez}},
  \bibinfo{author}{\bibfnamefont{D.}~\bibnamefont{Levesque}}, \bibnamefont{and}
  \bibinfo{author}{\bibfnamefont{J.-J.} \bibnamefont{Weis}},
  \bibinfo{journal}{Phys. Rev. E} \textbf{\bibinfo{volume}{60}},
  \bibinfo{pages}{5495} (\bibinfo{year}{1999}).

\bibitem[{\citenamefont{Sarkisov and Monson}(2000)}]{sarkisov.monson:2000}
\bibinfo{author}{\bibfnamefont{L.}~\bibnamefont{Sarkisov}} \bibnamefont{and}
  \bibinfo{author}{\bibfnamefont{P.~A.} \bibnamefont{Monson}},
  \bibinfo{journal}{Phys. Rev. E} \textbf{\bibinfo{volume}{61}},
  \bibinfo{pages}{7231} (\bibinfo{year}{2000}).

\bibitem[{\citenamefont{{De Grandis} et~al.}(2004)\citenamefont{{De Grandis},
  Gallo, and Rovere}}]{grandis.gallo.ea:2004}
\bibinfo{author}{\bibfnamefont{V.}~\bibnamefont{{De Grandis}}},
  \bibinfo{author}{\bibfnamefont{P.}~\bibnamefont{Gallo}}, \bibnamefont{and}
  \bibinfo{author}{\bibfnamefont{M.}~\bibnamefont{Rovere}},
  \bibinfo{journal}{Phys. Rev. E} \textbf{\bibinfo{volume}{70}},
  \bibinfo{pages}{061505} (\bibinfo{year}{2004}).

\bibitem[{\citenamefont{Sch{\"o}ll-Paschinger
  et~al.}(2001)\citenamefont{Sch{\"o}ll-Paschinger, Levesque, Weis, and
  Kahl}}]{scholl-paschinger.levesque.ea:2001}
\bibinfo{author}{\bibfnamefont{E.}~\bibnamefont{Sch{\"o}ll-Paschinger}},
  \bibinfo{author}{\bibfnamefont{D.}~\bibnamefont{Levesque}},
  \bibinfo{author}{\bibfnamefont{J.-J.} \bibnamefont{Weis}}, \bibnamefont{and}
  \bibinfo{author}{\bibfnamefont{G.}~\bibnamefont{Kahl}},
  \bibinfo{journal}{Phys. Rev. E} \textbf{\bibinfo{volume}{64}},
  \bibinfo{pages}{011502} (\bibinfo{year}{2001}).

\bibitem[{\citenamefont{Schmidt et~al.}(2002)\citenamefont{Schmidt,
  Sch{\"o}ll-Paschinger, K{\"o}finger, and
  Kahl}}]{schmidt.scholl-paschinger.ea:2002}
\bibinfo{author}{\bibfnamefont{M.}~\bibnamefont{Schmidt}},
  \bibinfo{author}{\bibfnamefont{E.}~\bibnamefont{Sch{\"o}ll-Paschinger}},
  \bibinfo{author}{\bibfnamefont{J.}~\bibnamefont{K{\"o}finger}},
  \bibnamefont{and} \bibinfo{author}{\bibfnamefont{G.}~\bibnamefont{Kahl}},
  \bibinfo{journal}{J. Phys.: Condens. Matter} \textbf{\bibinfo{volume}{14}},
  \bibinfo{pages}{12099} (\bibinfo{year}{2002}).

\bibitem[{\citenamefont{Kierlik et~al.}(1996)\citenamefont{Kierlik, Rosinberg,
  Tarjus, and Monson}}]{kierlik.rosinberg.ea:1996}
\bibinfo{author}{\bibfnamefont{E.}~\bibnamefont{Kierlik}},
  \bibinfo{author}{\bibfnamefont{M.~L.} \bibnamefont{Rosinberg}},
  \bibinfo{author}{\bibfnamefont{G.}~\bibnamefont{Tarjus}}, \bibnamefont{and}
  \bibinfo{author}{\bibfnamefont{P.~A.} \bibnamefont{Monson}},
  \bibinfo{journal}{J. Phys.: Condens. Matter} \textbf{\bibinfo{volume}{8}},
  \bibinfo{pages}{9621} (\bibinfo{year}{1996}).

\bibitem[{\citenamefont{Villain}(1982)}]{villain:1982}
\bibinfo{author}{\bibfnamefont{J.}~\bibnamefont{Villain}}, \bibinfo{journal}{J.
  Phys. (Paris)} \textbf{\bibinfo{volume}{43}}, \bibinfo{pages}{L551}
  (\bibinfo{year}{1982}).

\bibitem[{\citenamefont{Fisher}(1986)}]{fisher:1986}
\bibinfo{author}{\bibfnamefont{D.~S.} \bibnamefont{Fisher}},
  \bibinfo{journal}{Phys. Rev. Lett.} \textbf{\bibinfo{volume}{56}},
  \bibinfo{pages}{416} (\bibinfo{year}{1986}).

\bibitem[{\citenamefont{Gofman et~al.}(1993)\citenamefont{Gofman, Adler,
  Aharony, Harris, and Schwartz}}]{gofman.adler.ea:1993}
\bibinfo{author}{\bibfnamefont{M.}~\bibnamefont{Gofman}},
  \bibinfo{author}{\bibfnamefont{J.}~\bibnamefont{Adler}},
  \bibinfo{author}{\bibfnamefont{A.}~\bibnamefont{Aharony}},
  \bibinfo{author}{\bibfnamefont{A.~B.} \bibnamefont{Harris}},
  \bibnamefont{and} \bibinfo{author}{\bibfnamefont{M.}~\bibnamefont{Schwartz}},
  \bibinfo{journal}{Phys. Rev. Lett.} \textbf{\bibinfo{volume}{71}},
  \bibinfo{pages}{1569} (\bibinfo{year}{1993}).

\bibitem[{\citenamefont{Fisher and Zinn}(1998)}]{fisher.zinn:1998}
\bibinfo{author}{\bibfnamefont{M.~E.} \bibnamefont{Fisher}} \bibnamefont{and}
  \bibinfo{author}{\bibfnamefont{S.-Y.} \bibnamefont{Zinn}},
  \bibinfo{journal}{J. Phys. A: Math. Gen.} \textbf{\bibinfo{volume}{31}},
  \bibinfo{pages}{L629} (\bibinfo{year}{1998}).

\bibitem[{\citenamefont{Rieger}(1995)}]{rieger:1995}
\bibinfo{author}{\bibfnamefont{H.}~\bibnamefont{Rieger}},
  \bibinfo{journal}{Phys. Rev. B} \textbf{\bibinfo{volume}{52}},
  \bibinfo{pages}{6659} (\bibinfo{year}{1995}).

\bibitem[{\citenamefont{Newman and Barkema}(1996)}]{newman.barkema:1996}
\bibinfo{author}{\bibfnamefont{M.~E.~J.} \bibnamefont{Newman}}
  \bibnamefont{and} \bibinfo{author}{\bibfnamefont{G.~T.}
  \bibnamefont{Barkema}}, \bibinfo{journal}{Phys. Rev. E}
  \textbf{\bibinfo{volume}{53}}, \bibinfo{pages}{393} (\bibinfo{year}{1996}).

\bibitem[{\citenamefont{Asakura and Oosawa}(1954)}]{asakura.oosawa:1954}
\bibinfo{author}{\bibfnamefont{S.}~\bibnamefont{Asakura}} \bibnamefont{and}
  \bibinfo{author}{\bibfnamefont{F.}~\bibnamefont{Oosawa}},
  \bibinfo{journal}{J. Chem. Phys.} \textbf{\bibinfo{volume}{22}},
  \bibinfo{pages}{1255} (\bibinfo{year}{1954}).

\bibitem[{\citenamefont{Vrij}(1976)}]{vrij:1976}
\bibinfo{author}{\bibfnamefont{A.}~\bibnamefont{Vrij}}, \bibinfo{journal}{Pure
  Appl. Chem.} \textbf{\bibinfo{volume}{48}}, \bibinfo{pages}{471}
  (\bibinfo{year}{1976}).

\bibitem[{\citenamefont{Aarts et~al.}(2002)\citenamefont{Aarts, Tuinier, and
  Lekkerkerker}}]{aarts.tuinier.ea:2002}
\bibinfo{author}{\bibfnamefont{D.}~\bibnamefont{Aarts}},
  \bibinfo{author}{\bibfnamefont{R.}~\bibnamefont{Tuinier}}, \bibnamefont{and}
  \bibinfo{author}{\bibfnamefont{H.}~\bibnamefont{Lekkerkerker}},
  \bibinfo{journal}{J. Phys.: Condens. Matter} \textbf{\bibinfo{volume}{14}},
  \bibinfo{pages}{7551} (\bibinfo{year}{2002}).

\bibitem[{\citenamefont{Brader et~al.}(2002)\citenamefont{Brader, Evans,
  Schmidt, and L{\"o}wen}}]{brader.evans.ea:2002}
\bibinfo{author}{\bibfnamefont{J.~M.} \bibnamefont{Brader}},
  \bibinfo{author}{\bibfnamefont{R.}~\bibnamefont{Evans}},
  \bibinfo{author}{\bibfnamefont{M.}~\bibnamefont{Schmidt}}, \bibnamefont{and}
  \bibinfo{author}{\bibfnamefont{H.}~\bibnamefont{L{\"o}wen}},
  \bibinfo{journal}{J. Phys.: Condens. Matter} \textbf{\bibinfo{volume}{14}},
  \bibinfo{pages}{L1} (\bibinfo{year}{2002}).

\bibitem[{\citenamefont{Bergenholtz et~al.}(2003)\citenamefont{Bergenholtz,
  Poon, and Fuchs}}]{bergenholtz.poon.ea:2003}
\bibinfo{author}{\bibfnamefont{J.}~\bibnamefont{Bergenholtz}},
  \bibinfo{author}{\bibfnamefont{W.}~\bibnamefont{Poon}}, \bibnamefont{and}
  \bibinfo{author}{\bibfnamefont{M.}~\bibnamefont{Fuchs}},
  \bibinfo{journal}{Langmuir} \textbf{\bibinfo{volume}{19}},
  \bibinfo{pages}{4493} (\bibinfo{year}{2003}).

\bibitem[{\citenamefont{Lekkerkerker et~al.}(1992)\citenamefont{Lekkerkerker,
  Poon, Pusey, Stroobants, and Warren}}]{lekkerkerker.poon.ea:1992}
\bibinfo{author}{\bibfnamefont{H.}~\bibnamefont{Lekkerkerker}},
  \bibinfo{author}{\bibfnamefont{W.}~\bibnamefont{Poon}},
  \bibinfo{author}{\bibfnamefont{P.}~\bibnamefont{Pusey}},
  \bibinfo{author}{\bibfnamefont{A.}~\bibnamefont{Stroobants}},
  \bibnamefont{and} \bibinfo{author}{\bibfnamefont{P.}~\bibnamefont{Warren}},
  \bibinfo{journal}{Europhys. Lett.} \textbf{\bibinfo{volume}{20}},
  \bibinfo{pages}{559} (\bibinfo{year}{1992}).

\bibitem[{\citenamefont{Vink and Horbach}(2004)}]{vink.horbach:2004*1}
\bibinfo{author}{\bibfnamefont{R.~L.~C.} \bibnamefont{Vink}} \bibnamefont{and}
  \bibinfo{author}{\bibfnamefont{J.}~\bibnamefont{Horbach}},
  \bibinfo{journal}{J. Chem. Phys.} \textbf{\bibinfo{volume}{121}},
  \bibinfo{pages}{3253} (\bibinfo{year}{2004}).

\bibitem[{\citenamefont{Schwartz et~al.}(1993)\citenamefont{Schwartz, Villain,
  Shapir, and Nattermann}}]{physrevb.48.3095}
\bibinfo{author}{\bibfnamefont{M.}~\bibnamefont{Schwartz}},
  \bibinfo{author}{\bibfnamefont{J.}~\bibnamefont{Villain}},
  \bibinfo{author}{\bibfnamefont{Y.}~\bibnamefont{Shapir}}, \bibnamefont{and}
  \bibinfo{author}{\bibfnamefont{T.}~\bibnamefont{Nattermann}},
  \bibinfo{journal}{Phys. Rev. B} \textbf{\bibinfo{volume}{48}},
  \bibinfo{pages}{3095} (\bibinfo{year}{1993}).

\bibitem[{\citenamefont{Virnau and M{\"u}ller}(2004)}]{virnau.muller:2004}
\bibinfo{author}{\bibfnamefont{P.}~\bibnamefont{Virnau}} \bibnamefont{and}
  \bibinfo{author}{\bibfnamefont{M.}~\bibnamefont{M{\"u}ller}},
  \bibinfo{journal}{J. Chem. Phys.} \textbf{\bibinfo{volume}{120}},
  \bibinfo{pages}{10925} (\bibinfo{year}{2004}).

\bibitem[{\citenamefont{Binder}(1981)}]{binder:1981}
\bibinfo{author}{\bibfnamefont{K.}~\bibnamefont{Binder}}, \bibinfo{journal}{Z.
  Phys. B: Condens. Matter} \textbf{\bibinfo{volume}{43}}, \bibinfo{pages}{119}
  (\bibinfo{year}{1981}).

\bibitem[{\citenamefont{Eichhorn and Binder}(1995)}]{eichhorn.binder:1995}
\bibinfo{author}{\bibfnamefont{K.}~\bibnamefont{Eichhorn}} \bibnamefont{and}
  \bibinfo{author}{\bibfnamefont{K.}~\bibnamefont{Binder}},
  \bibinfo{journal}{Europhys. Lett.} \textbf{\bibinfo{volume}{30}},
  \bibinfo{pages}{331} (\bibinfo{year}{1995}).

\bibitem[{\citenamefont{Eichhorn and Binder}(1996)}]{eichhorn.binder:1996}
\bibinfo{author}{\bibfnamefont{K.}~\bibnamefont{Eichhorn}} \bibnamefont{and}
  \bibinfo{author}{\bibfnamefont{K.}~\bibnamefont{Binder}},
  \bibinfo{journal}{J. Phys.: Condens. Matter} \textbf{\bibinfo{volume}{8}},
  \bibinfo{pages}{5209} (\bibinfo{year}{1996}).

\bibitem[{\citenamefont{Fisher}(1974)}]{fisher:1974}
\bibinfo{author}{\bibfnamefont{M.~E.} \bibnamefont{Fisher}},
  \bibinfo{journal}{Rev. Mod. Phys.} \textbf{\bibinfo{volume}{46}},
  \bibinfo{pages}{597} (\bibinfo{year}{1974}).

\bibitem[{\citenamefont{Binder and Deutsch}(1992)}]{binder.deutsch:1992}
\bibinfo{author}{\bibfnamefont{K.}~\bibnamefont{Binder}} \bibnamefont{and}
  \bibinfo{author}{\bibfnamefont{H.-P.} \bibnamefont{Deutsch}},
  \bibinfo{journal}{Europhys. Lett.} \textbf{\bibinfo{volume}{18}},
  \bibinfo{pages}{667} (\bibinfo{year}{1992}).

\bibitem[{\citenamefont{Luijten et~al.}(2002)\citenamefont{Luijten, Fisher, and
  Panagiotopoulos}}]{luijten.fisher.ea:2002}
\bibinfo{author}{\bibfnamefont{E.}~\bibnamefont{Luijten}},
  \bibinfo{author}{\bibfnamefont{M.~E.} \bibnamefont{Fisher}},
  \bibnamefont{and} \bibinfo{author}{\bibfnamefont{A.~Z.}
  \bibnamefont{Panagiotopoulos}}, \bibinfo{journal}{Phys. Rev. Lett.}
  \textbf{\bibinfo{volume}{88}}, \bibinfo{pages}{185701}
  (\bibinfo{year}{2002}).

\bibitem[{\citenamefont{Kim et~al.}(2003)\citenamefont{Kim, Fisher, and
  Luijten}}]{kim.fisher.ea:2003}
\bibinfo{author}{\bibfnamefont{Y.~C.} \bibnamefont{Kim}},
  \bibinfo{author}{\bibfnamefont{M.~E.} \bibnamefont{Fisher}},
  \bibnamefont{and} \bibinfo{author}{\bibfnamefont{E.}~\bibnamefont{Luijten}},
  \bibinfo{journal}{Phys. Rev. Lett.} \textbf{\bibinfo{volume}{91}},
  \bibinfo{pages}{065701} (\bibinfo{year}{2003}).

\bibitem[{\citenamefont{Royall et~al.}(2006)\citenamefont{Royall, Aarts, and
  Tanaka}}]{royal.aarts.ea:2006}
\bibinfo{author}{\bibfnamefont{C.~P.} \bibnamefont{Royall}},
  \bibinfo{author}{\bibfnamefont{D.}~\bibnamefont{Aarts}}, \bibnamefont{and}
  \bibinfo{author}{\bibfnamefont{H.}~\bibnamefont{Tanaka}},
  \emph{\bibinfo{title}{Bridging lengthscales in colloidal liquid-vapour
  interfaces: from critical divergence to single particles}},
  \bibinfo{howpublished}{submitted} (\bibinfo{year}{2006}).

\bibitem[{\citenamefont{Kluijtmans et~al.}(2000)\citenamefont{Kluijtmans,
  Koenderink, and Philipse}}]{physreve.61.626}
\bibinfo{author}{\bibfnamefont{S.~G. J.~M.} \bibnamefont{Kluijtmans}},
  \bibinfo{author}{\bibfnamefont{G.~H.} \bibnamefont{Koenderink}},
  \bibnamefont{and} \bibinfo{author}{\bibfnamefont{A.~P.}
  \bibnamefont{Philipse}}, \bibinfo{journal}{Phys. Rev. E}
  \textbf{\bibinfo{volume}{61}}, \bibinfo{pages}{626} (\bibinfo{year}{2000}).

\bibitem[{\citenamefont{Kluijtmans and
  Philipse}(1999)}]{kluijtmans.philipse:1999}
\bibinfo{author}{\bibfnamefont{S.~G. J.~M.} \bibnamefont{Kluijtmans}}
  \bibnamefont{and} \bibinfo{author}{\bibfnamefont{A.~P.}
  \bibnamefont{Philipse}}, \bibinfo{journal}{Langmuir}
  \textbf{\bibinfo{volume}{15}}, \bibinfo{pages}{1896} (\bibinfo{year}{1999}).

\bibitem[{\citenamefont{Vossen et~al.}(2004)\citenamefont{Vossen, {van der
  Horst}, Dogterom, and van Blaaderen}}]{vossen.van-der-horst.ea:2004}
\bibinfo{author}{\bibfnamefont{D.~L.~J.} \bibnamefont{Vossen}},
  \bibinfo{author}{\bibfnamefont{A.}~\bibnamefont{{van der Horst}}},
  \bibinfo{author}{\bibfnamefont{M.}~\bibnamefont{Dogterom}}, \bibnamefont{and}
  \bibinfo{author}{\bibfnamefont{A.}~\bibnamefont{van Blaaderen}},
  \bibinfo{journal}{Review of Scientific Instruments}
  \textbf{\bibinfo{volume}{75}}, \bibinfo{pages}{2960} (\bibinfo{year}{2004}).

\bibitem[{\citenamefont{van Blaaderen}(1997)}]{blaaderen:1997}
\bibinfo{author}{\bibfnamefont{A.}~\bibnamefont{van Blaaderen}},
  \bibinfo{journal}{Progress in Colloid and Interface Science}
  \textbf{\bibinfo{volume}{104}}, \bibinfo{pages}{59} (\bibinfo{year}{1997}).

\bibitem[{\citenamefont{Aarts et~al.}(2004)\citenamefont{Aarts, Schmidt, and
  Lekkerkerker}}]{aarts.schmidt.ea:2004}
\bibinfo{author}{\bibfnamefont{D.~G. A.~L.} \bibnamefont{Aarts}},
  \bibinfo{author}{\bibfnamefont{M.}~\bibnamefont{Schmidt}}, \bibnamefont{and}
  \bibinfo{author}{\bibfnamefont{H.~N.~W.} \bibnamefont{Lekkerkerker}},
  \bibinfo{journal}{Science} \textbf{\bibinfo{volume}{304}},
  \bibinfo{pages}{847} (\bibinfo{year}{2004}).

\end{thebibliography}

\end{document}